\documentclass[12pt,preprint]{aastex}
\usepackage{graphicx}
\usepackage{amssymb}
\newcommand{\chandra}{{\it CHANDRA}}
\newcommand{\rxte}{{\it RXTE}}

\newcommand{\ec}{$\eta$ Carinae}
\slugcomment{Submitted to the Astrophysical Journal}

\shorttitle{X-ray Emission from the Homunculus Nebula}
\shortauthors{Corcoran et al.}

\begin{document}

%% LaTeX will automatically break titles if they run longer than
%% one line. However, you may use \\ to force a line break if
%% you desire.

\title{Waiting in the Wings: Reflected X-ray Emission from the Homunculus Nebula}

%% Use \author, \affil, and the \and command to format
%% author and affiliation information.
%% Note that \email has replaced the old \authoremail command
%% from AASTeX v4.0. You can use \email to mark an email address
%% anywhere in the paper, not just in the front matter.
%% As in the title, you can use \\ to force line breaks.

\author{M. F. Corcoran\altaffilmark{1,2}, 
K. Hamaguchi\altaffilmark{2}, 
T. Gull\altaffilmark{3},
K. Davidson\altaffilmark{4},
R. Petre\altaffilmark{2},
D. J. Hillier\altaffilmark{5},
N. Smith\altaffilmark{6},
A. Damineli\altaffilmark{7},
J. A. Morse\altaffilmark{8},
N. R. Walborn\altaffilmark{9},
E. Verner\altaffilmark{10},
N. Collins\altaffilmark{10},
S. White\altaffilmark{11},
J. M. Pittard\altaffilmark{12},
K. Weis\altaffilmark{13},
D. Bomans\altaffilmark{13},
Y. Butt\altaffilmark{14}}

%\and

%\author{R. J. Hanisch\altaffilmark{5}}
%\affil{Space Telescope Science Institute, Baltimore, MD 21218}

%% Notice that each of these authors has alternate affiliations, which
%% are identified by the \altaffilmark after each name.  Specify alternate
%% affiliation information with \altaffiltext, with one command per each
%% affiliation.

\altaffiltext{1}{Universities Space Research Association, 7501 Forbes 
Blvd, Ste 206, Seabrook, MD 20706} 
\altaffiltext{2}{Laboratory for High Energy Astrophysics, Goddard Space Flight 
Center, Greenbelt MD 20771; corcoran@lheapop.gsfc.nasa.gov; kenji@lheapop.gsfc.nasa.gov} 
\altaffiltext{3}{Laboratory for Astronomy and Solar Physics, NASA/Goddard Space Flight Center, Code 681, Greenbelt, MD 20771 USA; gull@sea.gsfc.nasa.gov}
\altaffiltext{4}{Astronomy Department, University of Minnesota, 116 Church St. SE, Minneapolis, MN 55455 US; kd@astro.umn.edu}
\altaffiltext{5}{Dept. of Physics and Astronomy, University of Pittsburgh, 3941 O'Hara St., Pittsburgh, PA 15260 USA; JDH@galah.phyast.pitt.edu}
\altaffiltext{6}{Center for Astrophysics and Space Astronomy, University of Colorado, 389 UCB, Boulder, CO 80309 USA; nathans@ozma.colorado.edu}
\altaffiltext{7}{Instituto de Astronomia, Geof\'{i}sica e Ci\^{e}ncias Atmosf\'{e}ricas, Universidade de S\~{a}o Paulo, Rua do Mat\~{a}o 1226, 05508-900 S\~{a}o Paulo, SP, Brazil; damineli@astro.iag.usp.br}
\altaffiltext{8}{Arizona State University, Department of Physics \& Astronomy, Box 871504, Tempe, AZ, 85287-1504; jon.Morse@asu.edu}
\altaffiltext{9}{Space Telescope Science Institute, 3700 San Martin Drive, Baltimore, MD 21218 USA; walborn@stsci.edu}
\altaffiltext{10}{IACS/Department of Physics, Catholic University of America, 620 Michigan Ave., NE Washington, DC 20064; kverner@fe2.gsfc.nasa.gov; collins@zolo.gsfc.nasa.gov}
\altaffiltext{11}{Department of Astronomy, University of Maryland, College Park, MD 20742 USA; white@astro.umd.edu}
\altaffiltext{12}{School of Physics and Astronomy, The University of Leeds, Woodhouse Lane, Leeds, LS2 9JT, UK; jmp@ast.leeds.ac.uk}
\altaffiltext{13}{Astronomisches Institut, Ruhr-Universit\"t{a}t Bochum, Universit\"atsstr. 150, 44780 Bochum, Germany; kweis@astro.rub.de; bomans@astro.ruhr-uni-bochum.de}
\altaffiltext{14}{Harvard-Smithsonian Center for Astrophysics, 60 Garden Street, Cambridge, MA 02138 USA; ybutt@head-cfa.harvard.edu}

%\altaffiltext{3}{present address: Center for Astrophysics,
%    60 Garden Street, Cambridge, MA 02138}
%\altaffiltext{4}{Visiting Programmer, Space Telescope Science Institute}
%\altaffiltext{5}{Patron, Alonso's Bar and Grill}

%% Mark off your abstract in the ``abstract'' environment. In the manuscript
%% style, abstract will output a Received/Accepted line after the
%% title and affiliation information. No date will appear since the author
%% does not have this information. The dates will be filled in by the
%% editorial office after submission.

\begin{abstract}

We report the first detection of X-ray emission associated with the Homunculus Nebula which surrounds the supermassive star \ec.  The emission is characterized by a temperature in excess of 100 MK, and is consistent with scattering of the time-delayed X-ray flux associated with the star.  The nebular emission is bright in the northwestern lobe and near the central regions of the Homunculus, and fainter in the southeastern lobe.  We also report the detection of an unusually broad Fe K fluorescent line, which may indicate fluorescent scattering off the wind of a companion star or some other high velocity outflow. The X-ray Homunculus is the nearest member of the small class of  Galactic X-ray reflection nebulae, and the only one in which both the emitting and reflecting sources are distinguishable. 

\end{abstract}

%% Keywords should appear after the \end{abstract} command. The uncommented
%% example has been keyed in ApJ style. See the instructions to authors
%% for the journal to which you are submitting your paper to determine
%% what keyword punctuation is appropriate.


%% From the front matter, we move on to the body of the paper.
%% In the first two sections, notice the use of the natbib \citep
%% and \citet commands to identify citations.  The citations are
%% tied to the reference list via symbolic KEYs. The KEY corresponds
%% to the KEY in the \bibitem in the reference list below. We have
%% chosen the first three characters of the first author's name plus
%% the last two numeral of the year of publication as our KEY for
%% each reference.

\section{Introduction}
The Homunculus  \citep{gav50} is a young, hollow, expanding bipolar nebula surrounding the extremely luminous and massive star \ec. It is believed to have a total mass of 2--12$M_{\odot}$ \citep{sgk,ns03a} and was ejected from the star during the ``Great Eruption'' of \ec\ in the 1840Õs.    At about 100$M_{\odot}$ \citep{dh82}  \ec\  is one of the most massive stars known. It possesses an enormous stellar wind, characterized by a mass loss rate  of  approximately $10^{-4} -10^{-3}M_{\odot}$ yr$^{-1}$, and a wind velocity of about 500 km s$^{-1}$ \citep{djh01}.  \ec\ is a strong source of hard X-rays \citep{fds79} and it undergoes a deep X-ray minimum every 5.53 years \citep{bish99, mfc01} that lasts for about 3 months.  It has been suggested \citep{corc97, jmp98, bish99, mfc00}  that the variable X-ray emission is produced by the collision of \ec's wind with a less dense, faster moving wind from an otherwise hidden companion (though alternative single-star models have been suggested, Davidson 1999). 
%In one binary star scenario the X-ray and far UV minimum occur when the companion is hidden behind \ecÕs massive wind at a time when the two stars happen to be at their closest approach.  In this scenario the intrinsic X-ray emission from the source itself is actually near its peak brightness during the observed X-ray minimum since the density of the shocked region is at its maximum when the stars are close, but this emission is almost totally blocked by the intervening absorption from \ecÕs massive wind.  
 \ec\ began its most recent X-ray minimum on June 29, 2003 \citep{mfc03} and lasted through September 3, 2003.
 
A new observation by the \chandra\ X-ray Observatory  during \ec's recent X-ray intensity minimum  has for the first time identified X-ray emission from the Homunculus nebula itself.  This emission is at least a factor of four fainter than the direct stellar emission detected during the minimum, and about a factor of 100 fainter than the direct stellar emission outside of the minimum.  The emission associated with the Homunculus was not seen previously by \chandra\ or other X-ray observatories since it is hidden by the wings of the instrumentally broadened central source outside of the X-ray minimum.  In  this paper we report the spectral, spatial and temporal properties of the X-rays from the Homunculus.  A more detailed discussion of these data regarding the X-ray emission from \ec\ and its surroundings during the X-ray eclipse interval will be presented elsewhere (Corcoran et al.. 2004a, in preparation).

This paper is organized as follows.  The observations are presented in \S \ref{obs}. The X-ray and optical imaging of \ec\ and its surroundings are compared in \S \ref{imaging}.  In \S \ref{spectrum} we analyze the X-ray spectrum of the Homunculus, and interpret this emission in \S \ref{interp}.  We discuss spatial and temporal variations of the emission in \S \ref{var}, and present our conclusions in \S \ref{conclusions}.   

\section{Observations}
\label{obs}

To help characterize the nature of the X-ray minimum, we planned five observations with the \chandra\ X-ray Observatory   before, during and after this event as part of a large, multi-wavelength observing campaign during the summer of 2003.  The \chandra\ observations used the Advanced CCD Imaging Spectrometer  (ACIS) spectroscopic array \citep{gar03} plus the High Energy Transmission Gratings \citep{mark94}.  One  observation was obtained on July 20, 2003 (\chandra\ sequence number  200216) when contemporaneous monitoring with the \rxte\ satellite observatory showed \ec\ to be in the middle of its X-ray eclipse\footnote{http://lheawww.gsfc.nasa.gov/users/corcoran/eta\_car/etacar\_rxte\_lightcurve/.}. The total exposure time was 90,275 seconds. We re-extracted good ``Level 2'' events from the ``Level 1'' events file using the \chandra\ Interactive Analysis of Observations (CIAO) software package and the processing steps recommended by the \chandra\ X-ray Center\footnote{http://cxc.harvard.edu/ciao/threads/data.html}.  We included corrections for charge transfer inefficiency (CTI) and destreaking. The effective resolution of the screened photon events file was increased using the algorithm of \citet{ts01}\footnote{\url{http://cxc.harvard.edu/cont-soft/software/subpixel\_resolution.1.4.html}}.

A followup director's discretionary time observation with ACIS was obtained on August 28, 2003 (sequence number 200237) in order to confirm the detection of X-ray emission from the Homunculus and to look for changes in the extended emission which might have occurred as the central source brightened. This observation was obtained with \ec\ placed on the ACIS S3 chip, though to minimize the exposure time required the gratings were not used.  In order to mitigate effects from pileup of the central source we used a $1/8$ sub-array during the observation.  The total exposure time of this observation was 18,796 seconds. As above, we re-extracted the level 2 photon events and corrected for CTI, destreaking and improved the spatial resolution using the algorithm of \citet{ts01}.

\section{X-ray and Optical Imaging}
\label{imaging}

\begin{figure}
\centerline {
\plotone{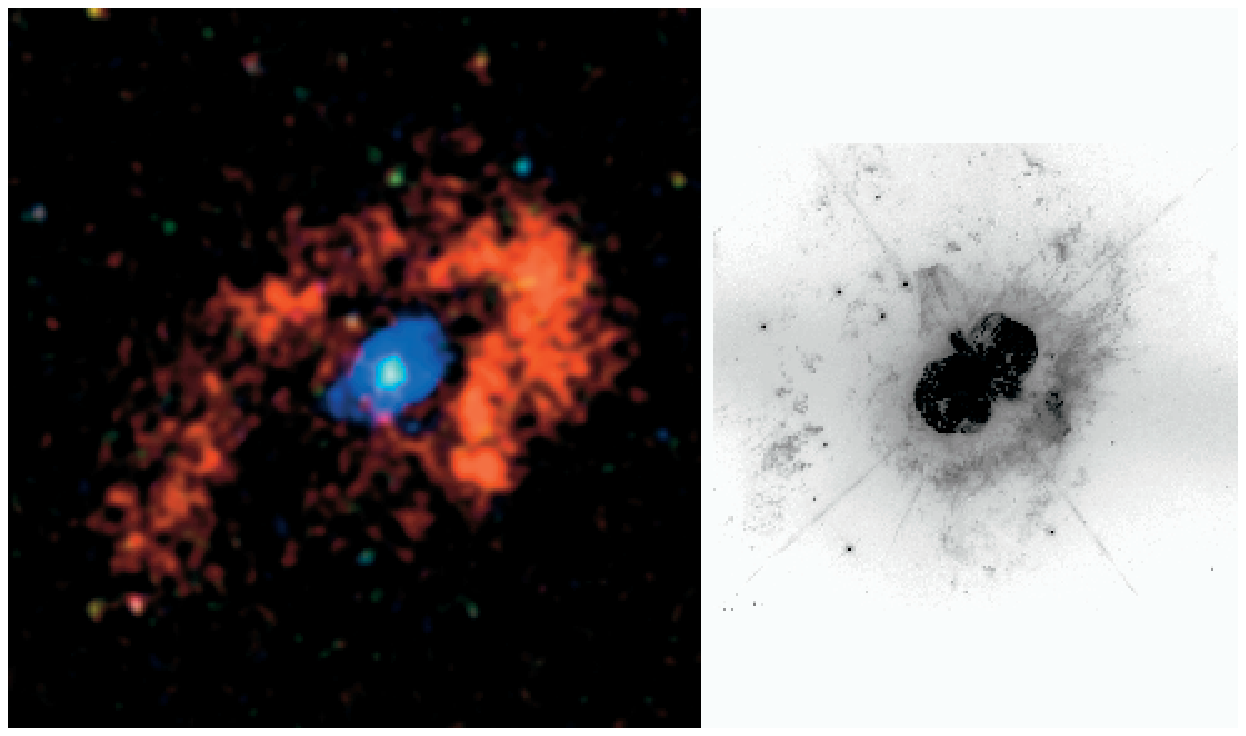}
}
\caption{\textit{Left}: A color X-ray image of \ec\ during the star's most recent X-ray minimum obtained by the \chandra\ X-ray observatory.  North is to the top, and East to the left. Red indicates low-energy X-ray emission ($0.2-1.5$ keV), green medium energy ($1.5-3.0$ keV) and blue high-energy emission ($3.0-8.0$ keV). Emission from the star itself is visible as a white point source at the center of the image.  A red, broken elliptical ring of emission lies beyond the star.  The bluish patch around the star inside this ring is reflected X-ray emission from the Homunculus Nebula.  \textit{Right}: An HST/WFPC2  [NII] $\lambda$6583 image of \ec.  The plate scale and orientation of the Hubble and \chandra\ images are the same.}
\label{image}
\end{figure}

Figure \ref{image} shows the zeroth-order X-ray image of \ec\ obtained by the ACIS-S3 CCD.  The color coding of the image represents emission from gas at different temperatures: red corresponds to the temperature range 2--15 million degrees (0.2--1.5 keV), green 15--30 million degrees (1.5--3.0 keV), and blue 30--100 million degrees (3--10 keV).  \ec\ is visible as a blue-white source at the center of the image. As seen previously \citep{fds01}, an incomplete elliptical ring or shell of soft X-ray emission (which appears red in figure \ref{image}) nearly surrounds \ec\ at a distance of about 2.2 light years from the star.  An elongated patch of bluish (hard X-ray) emission about $17''$  in length (about 0.6 light years) is clearly visible in Figure \ref{image} between the star and the outer Òred shellÓ. This hard emission was probably detected in  earlier ACIS observations \citep{fds01, kw04}, as a halo of scattered X-rays surrounding the central stellar source; however contamination by X-rays from the bright, heavily piled up central source did not allow this halo to be clearly identified. Figure \ref{image} also shows an HST/WFPC2 image of the nebulae surrounding \ec, at the same scale and orientation as the X-ray image. \ec\ is not visible in the image, which is scaled in intensity to highlight the nebulosities.  The hard ``blue'' emission in the X-ray image and the optical image of the Homunculus show marked similarities in size, shape and orientation.

\begin{figure}
\centerline {
\includegraphics[width=5in]{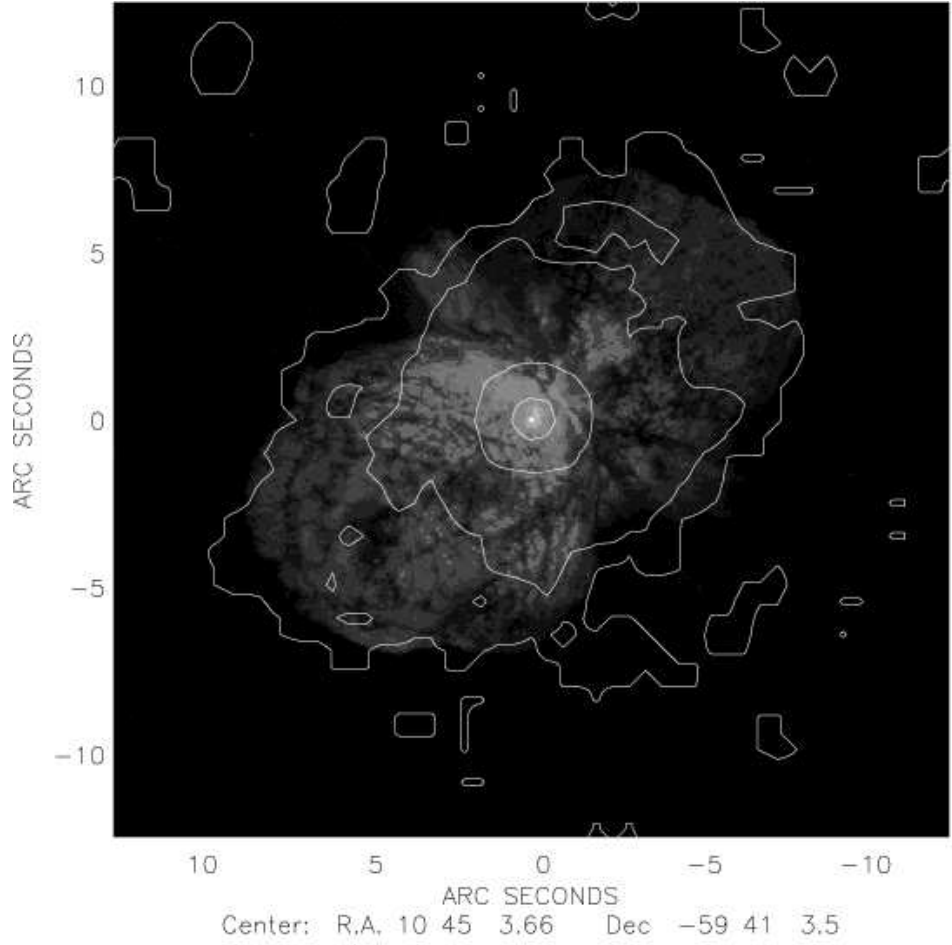}
}
\caption{Contours of the $3-8$ keV X-ray emission on a Hubble ACS image of the Homunculus.  Contour levels are  0.1, 0.7, 6.9  and  68.9\% of the peak brightness.}
\label{contour}
\end{figure}

\begin{figure}
\centerline {\includegraphics[width=6in]{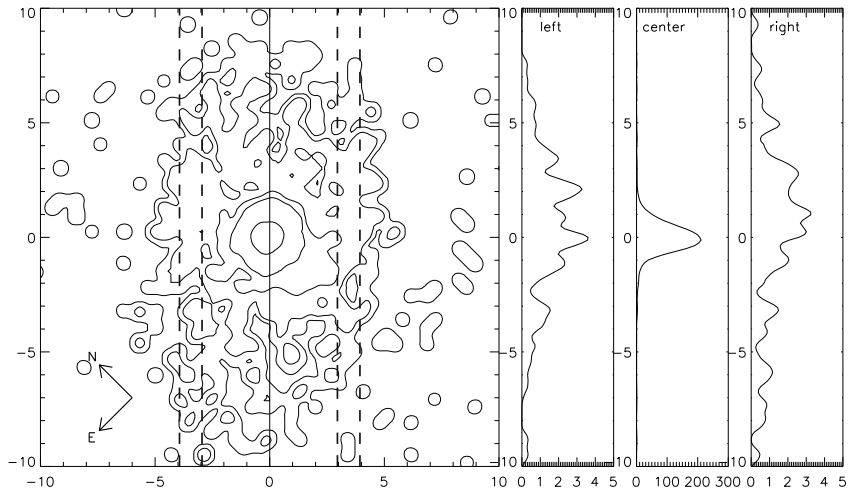}}
\caption{The image on the left is the contour map of the 3--8 keV X-ray surface brightness of the Homunculus. Surface brightness profiles $10''$ to the left of the star, on the star, and $10''$ to the right of the star are shown next to the contoured image.  The location of the surface brightness cuts on either side of the star is shown by the vertical dashed lines. The left and right brightness profiles are integrated between the dashed lines.}
\label{sb}
\end{figure}

Figure \ref{contour} shows an overlay of the 3--8 keV band X-ray contours on an HST image obtained by the Advanced Camera for Surveys \citep{ford03} as part of the HST \ec\ ``Treasury Project''\footnote{\url{http://etacar.umn.edu/}}.  The isophots of the hard emission clearly  follow the shape of the Homunculus, and  there is very little hard X-ray emission beyond the Homunculus. Figure \ref{sb} shows the spatial variation of X-ray surface brightness through the Homunculus for 3 impact parameters, one through the central star and others between 3 and 4 arcseconds on either side of the star.  The X-ray brightness is not uniform: the northwest lobe, and regions near the star are apparently brighter than the southeast lobe. 

\section{X-ray Spectroscopy of the Homunculus}
\label{spectrum}
 We extracted an X-ray spectrum of the Homunculus from an elliptical region of $22'' \times 14''$ oriented with the semi-major axis along the polar axis of the Homunculus,  excluding emission within 2.5 arcseconds of the star (which should exclude more than 95\% of the direct stellar flux) and soft emission from the outer ejecta.   The X-ray spectrum of the Homunculus is shown in Figure \ref{spec}.  This spectrum has been corrected for background emission, which was estimated from a source-free region to the northwest of the ``red shell''.   
 
 \subsection{The Thermal Component}
 We created responses and effective areas for the ACIS observation, and fit the extracted spectrum with a combination of absorbed collisionally-ionized plasma models using the ``mekal'' model within  the XSPEC analysis package \citep{xspec}. For simplicity we assumed solar abundances although the composition of material around \ec\ is decidedly non-solar.  The X-ray emission associated with the Homunculus appears to be dominated by a component with a characteristic temperature of 113 million degrees, along with a weaker, cooler component ($\sim7$ million degrees).  Table 1 compares the spectral properties (temperature, column density and luminosities) for the cooler and hotter components of the Homunculus, and the X-ray spectrum extracted from a $3.5''$ circle around the central star (corrected for background using the same background region as above). 

The hard emission from the Homunculus is strongly absorbed, having an equivalent hydrogen column density ($N_{H}) = 1.5 \times 10^{23}$ atoms cm$^{-2}$, which is about the same as the absorbing column to the star.  The absorbing column to the 6.7 million degree  component is only $N_{H}\approx2\times10^{21}$ atoms cm$^{-2}$, which is consistent with foreground absorption.  An examination of a low-energy X-ray image shows that this emission is clearly associated with the  X-ray ``bridge'' \citep{kw02, kw04} running across the middle of the Homunculus.  

\subsection{Fluorescent Emission}
\begin{figure}
\centerline {
\plotone{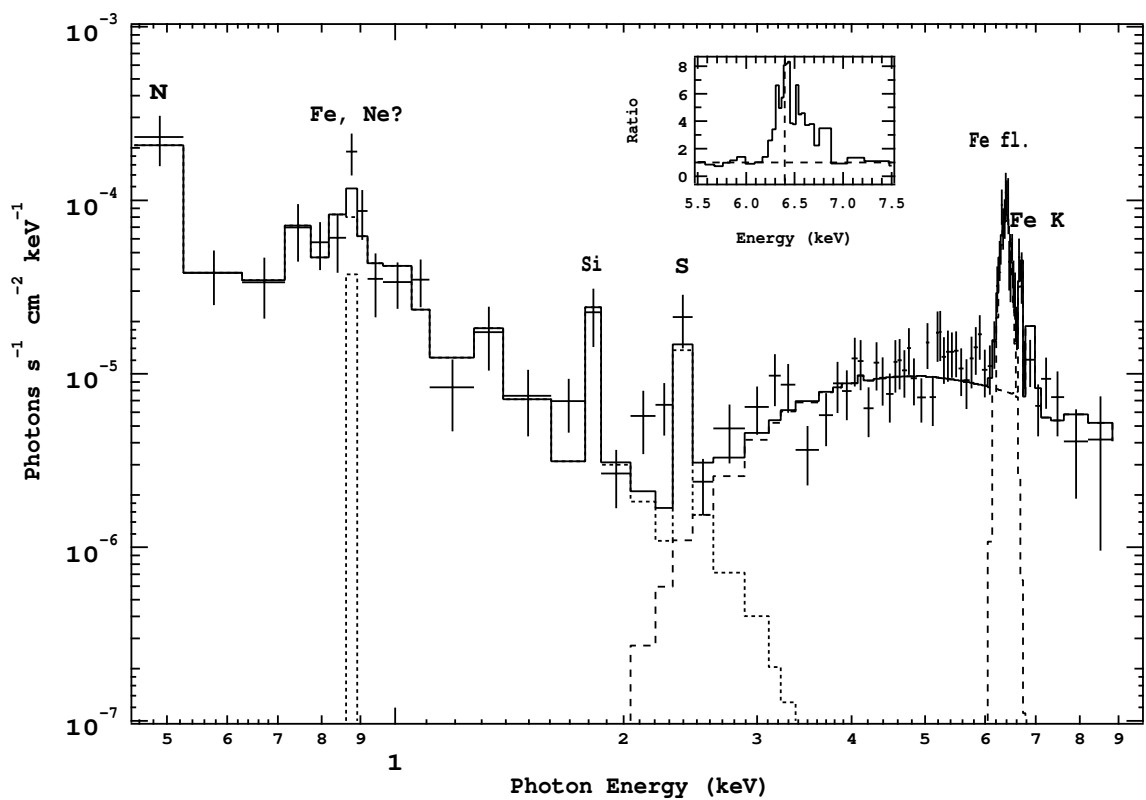}
}
\caption{The X-ray spectrum of the Homunculus from the July 20th observation.  The model spectrum is shown as an unbroken line, with individual components shown by broken lines.  An emission line near 0.9 keV may be emission from Fe XVII or Ne IX, while the strongest emission line at 6.4 keV is produced by fluorescent scattering of X-rays by iron atoms in the Homunculus.  Emission from at least two plasmas, one at a temperature near 7 million degrees and one at a temperature of about 100 million degrees  are required to describe the observed spectrum.  The inset shows the ratio of the emission near the Fe line complex at 6.4-6.8 keV to a simple power law model to describe the continuum emission.  Significant emission below 6.4 keV is observed in the Fe K fluorescent line.}
\label{spec}
\end{figure}

The strongest feature in the X-ray spectrum of the Homunculus  is an iron K-shell line at 6.4 keV, which we have modelled as a Gaussian in Figure \ref{spec}. The inset in Figure \ref{spec} shows the ratio of the observed emission to a simple power law model which describes the continuum emission near the Fe line complex.   The 6.4 keV line is produced by fluorescent scattering of X-rays by iron atoms (in either  a gaseous phase inside the Homunculus, or locked in solid dust grains).  The equivalent width of this iron line is about 1.5 keV, which is about a factor of 5 larger than the equivalent width of the iron fluorescent line seen in the spectrum of the star at times outside of the X-ray minimum.  The line is significantly broadened ($\sigma\approx0.1$ keV) corresponding to a Doppler width of about 4700 km s$^{-1}$. The fluorescent line in the stellar spectrum outside of minimum is consistent with $\sigma=0$ \citep{mfc01b}.  The velocity width of the line measured here is much larger than twice the measured expansion velocity of the Homunculus, which is only about 650 km s$^{-1}$.  The line broadening might be indicative of scattering off higher velocity flows in the lobes of the Homunculus.  The velocity of \ec's wind near the poles of the star has been measured to be $\approx 1100$ km s$^{-1}$ \citep{ns03b}, so that scattering from this material would produce a line broadening of $\lesssim 2200$ km s$^{-1}$, still a factor of 2 or more too low.   At energies $> 6.4$ kev, some of the observed broadening is probably due to blending of narrow fluorescent Fe lines from Fe in a range of ionization stages.  However, the fluorescent line shows an apparent ``red wing'' (as seen in the inset in figure \ref{spec}) extending to  $\sim6.2$ keV, which cannot be produced from line blending.  Perhaps processes more exotic than simple Doppler broadening are in evidence. For example, Compton scattering can produce a ``core'' line with low energy ``shoulders'' \citep{matt02} though the intensity of the shoulder is expected to be much lower than the intensity of the core, which does not appear to be the case here (though it is conceivable that the line profile is produced by some superposition core lines + Compton shoulders given the complex scattering surfaces in the Homunculus).  On the other hand, in systems like low-mass X-ray binaries and active galaxies, the presence of a compact object can give rise to  broad Fe K lines with significant red wings due to gravitational redshifts in accretion disks, but there is no strong evidence for the presence of a compact object in the \ec\ system.  It is perhaps most likely that the red wing is produced by X-ray scattering off neutral iron in the wind of \ec's companion star, since the companion's wind  velocity is believed to be $\sim 3000$ km s$^{-1}$ \citep{pc02}, and the companion's wind should be receding from the observer at the time of the \chandra\ observation, based on the orbit presented in \citet{mfc01}.  However, the companion's wind should be confined near the orbital plane, while the broad fluorescent line is apparently distributed throughout the entire Homunculus.
% Including 2 narrow lines instead of a single broad line results in line centroids of $6.31$ and $6.49$ keV. 
%lobes_minus_etabin10_2gauss.xcm
%An iron fluorescent line image shows that the fluorescing material is distributed throughout the Homunculus.  

\section{Interpretation of the X-ray Emission from the Homunculus}
\label{interp}

The high energy X-ray spectrum  of the Homunculus is similar to the hard X-ray emission from \ec\ itself  in temperature and column density. Such emission could arise from shock-heated gas in the lobes of the Homunculus, but while shocked gas has been observed in  \citep{ns02} and around \citep{dc03} the lobes, the deduced velocities are too low to produce significant emission at 8 keV.  Furthermore, high velocity shocked gas could not exist in the lobes since such shocks are inconsistent with the presence of  H$_{2}$ which is seen in the lobes \citep{ns02}. The high energy X-rays we observe  from the Homunculus are instead an X-ray ``light echo'' in which hard X-ray photons produced in a strong, high-velocity shock near  \ec\ are reflected to the observer at earth, primarily by Thomson scattering off electrons in the Homunculus.  Although the Homunculus contains significant amounts of dust, scattering from dust grains could not produce the Homunculus X-rays since dust scattering is so strongly forward peaked \citep{ps95}.

The amount of scattering depends approximately on the ratio of the scattering cross section to the mass-weighted geometrical cross section of the lobes.  
For simplicity we consider the Homunculus as 2 hollow connected spheres with an axis running from the point of contact of the spheres through their centers, and assume that the axis is tilted by $45^{\circ}$ to the line of sight, with the northwest lobe tilted away from the observer.  Assuming that the composition of the Homunculus is
50\%  helium and 50\% hydrogen, the total number of electrons is roughly $N_{e}<5 \times 10^{57}$ and the total Thomson-scattering cross section is $A_{e}=N_{e} \times  \sigma_{T} < 3.6 \times 10^{33}$ cm$^{2}$, where $\sigma_{T}$ is the Thomson scattering cross section ($0.665\times10^{-24}$ cm$^{2}$), assuming each lobe contains 6 solar masses \citep{ns03a} as an upper limit.  The approximate geometric radius of a lobe is very roughly $R_{L} = 2.3 \times 10^{17}$ cm after weighting the lobe by the derived mass distribution \citep{kd01}.  Thus the probability of scattering for an X-ray photon is roughly $P_{scat} < (A_{e}/4\pi R_{L}^{2}) =  0.5$\%. 
The observed scattering probability is roughly $P_{scat} \approx I_{H}(t)/I_{\eta}(t')$, where $I_{H}(t)$ is the intensity of the emission reflected from the Homunculus at the time of the \chandra\ observation, and $I_{\eta}(t')$ is the intensity of the star at an earlier time $t'=t-\Delta t$, where $\Delta t$ is the light-travel time between the star and the reflecting surface, $\Delta t\approx R_{L}/c \approx 88$ days.  Monitoring with the \rxte\ observatory shows, 88 days prior to the \chandra\ observation (April 23, 2003, observation identification 80001-01-54-00, Corcoran et al. 2004b, in preparation),  that $I_{\eta}(t') \approx  1.6\times 10^{35}$ 
ergs s$^{-1}$, so that the observed scattering probability $P_{scat} \approx 3.4\times10^{31}/1.6\times10^{35}=0.2$\%, in very good agreement with the derivation given above for regions with appreciable light travel times.  

\section{Spatial and Temporal Variations}
\label{var}

Regions of  the Homunculus near the star where light travel times are negligible are also X-ray bright.  Figure \ref{sb} shows that the X-ray emission in the Homunculus on either side of the star is noticeably peaked near the star, though there is much less contrast in surface brightness on either side of the star.   This suggests that, in the direction of the low-latitude reflecting surfaces, the stellar source may be substantially brighter than the stellar flux that we see directly, and/or the amount of scattering material in the region near the star may be much higher than the amount of material in the lobes.  Another alternative is that there may be substantially more extinction to the X-ray emitting region along our line of sight than in the direction of these reflecting surfaces.  In principal some or all of these suggestions may play a role.  Detailed analysis of the variation of the surface brightness profiles depends on the time-varying flux from the central source, the real distribution of matter in the lobes, and the amount of material between the star and the reflecting surfaces.

\begin{figure}
\centerline {
\includegraphics[angle=90,width=6in]{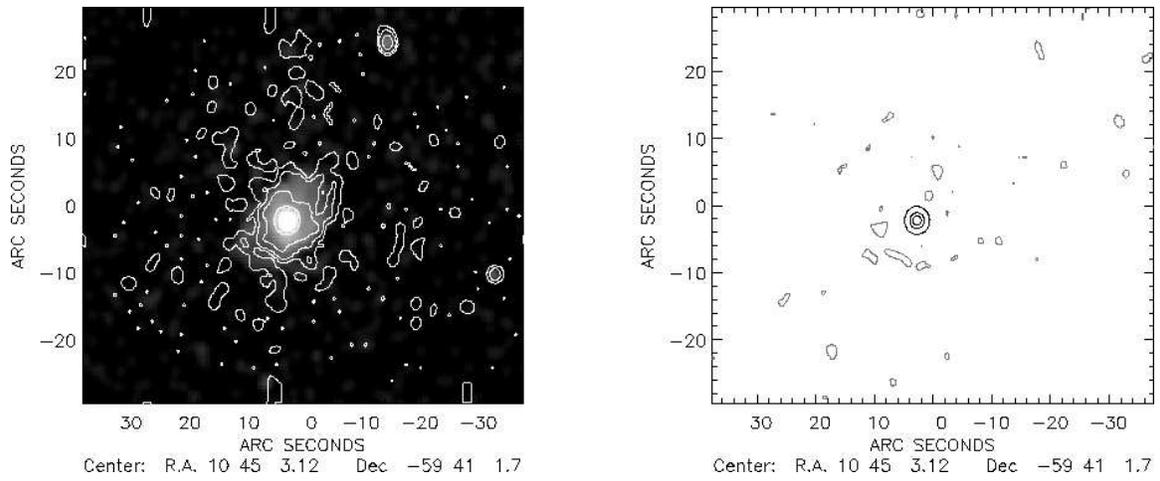}
}
\caption{The image on the left shows contours of the August 28 2003 observation on the July 20th ACIS zeroth-order image.  The image on the right shows a contour image of the exposure corrected August 28 image minus the exposure-corrected July 20 image.  Negative contours are shown in gray.  The central source is significantly brighter in August 28 compared to the earlier observation.}
\label{cmp}
\end{figure}

Figure \ref{cmp} shows a comparison between the ACIS S3 image obtained on August 28 and the HETG+ACIS S3 zeroth order image from the July 20th observation.  The images have been exposure corrected.  The image on the left shows the August 28 image contours on the July 20th image. Both images are in the 3--8 keV band.  The shape and intensity of the emission from the Homunculus is very similar in the two observations.  Figure \ref{cmp} also shows a difference image in which the exposure corrected August 28 image is subtracted from the exposure corrected July 20 image.  Contours in gray are negative contours (July 20 brighter than August 28) and black are positive.  Though the brightness of the central source has increased substantially between the two observations, there is little change in intensity in the reflected emission from the Homunculus, although there is perhaps some evidence of a fading in the southeastern lobe.  

\section{Conclusions}
\label{conclusions}

We have presented the first clear detection of X-ray emission from the Homunculus nebula around \ec, and have shown that this emission is consistent with the amount of stellar emission expected to be scattered from the Homunculus accounting for light-travel-time delays.  Along with the scattered emission, we note the presence of strong Fe K fluorescent line at 6.4 keV, which is anomalously broad, and which shows emission below 6.4 keV.

The X-ray Homunculus is the nearest member of the small class of  Galactic X-ray reflection nebulae \citep{mur03}, and the only one where both the emitting source and reflector are clearly identified. Variation in the intensity, orientation and spectral shape of the central  source will be mirrored by variations in the scattered emission from the Homunculus, and in principle can 
help define the three-dimensional structure by providing view of the stellar X-ray source from multiple lines of sight, similar to the way in which reflected optical emission provides a latitudinally-dependent view of \ec\ itself \citep{ns03b}.   Continued monitoring of the reflected emission after the star's X-ray minimum would be especially interesting, since this could provide a measure of the intrinsic X-ray luminosity of the star during the minimum, and could help distinguish an occultation of the X-ray source from variation in the source's intrinsic X-ray brightness.   Additional high spatial resolution observations of the X-ray echo from the Homunculus can help define the three-dimensional structure and variability of the stellar X-ray source.  Observations of the reflected emission can also help map out the geometry of the absorbing material between the emitting region and the reflecting surface and allow us to more accurately model this unique system.

\acknowledgments

We gratefully acknowledge the exceptional support of  Dr. Fred Seward and Dr. Norbert Schulz of the \chandra\ X-ray Center for their help in scheduling these observations.  We also would like to thank the anonymous referee for helpful comments.  This work was supported by SAO grant \#GO3-4008A. This research has made use of NASA's Astrophysics Data System.  
This research has made use of software obtained from the High Energy Astrophysics Science Archive Research Center (HEASARC), provided by NASA's Goddard Space Flight Center, software developed and provided by the \chandra\ X-ray Center, and the subpixel\_resolution software developed by K. Mori at Penn State.

\clearpage

%% Use the figure environment and \plotone or \plottwo to include 
%% figures and captions in your electronic submission.

\vspace{.2in}
\vspace{.2in}

\clearpage 

\vspace{.2in}
\vspace{.2in}
\clearpage

%\vspace{.2in}
%\vspace{.2in}

\clearpage 

\vspace{.2in}
\vspace{.2in}

\clearpage 

\vspace{-2.0in}
%\vspace{.2in}

\clearpage

\begin{table}
\begin{center}
\caption{Spectral Characteristics of the Emission from the Homunculus\label{tbl1}}
\begin{tabular}{lcccc}
\tableline\tableline
Component & Temperature & Column Density &$L_{x}$\tablenotemark{a} & $L_{x,c}$\tablenotemark{b} \\
 & ($10^{6}$ K) & ($10^{22}$ cm$^{-2}$) & ($10^{31}$ ergs s$^{-1}$) &  ($10^{31}$ ergs s$^{-1}$)  \\
\tableline
Homunculus (soft)  & $6.7\pm2.7$   & $0.2\pm0.1$ & $17.7\pm2.1$ & 19.8 \\
Homunculus (hard) & $113\pm66$   & $15\pm3.3$  & $46.6\pm4.3$ & 113\\
star                         & $58.0\pm3.5$ & $15\pm4.5$  & $142.4\pm5.1$ & 311.4 \\
\tableline
\end{tabular}

% for errors, see table_errors.txt

\tablenotetext{a}{Observed luminosity in the 0.5--10 keV band, assuming the distance to \ec\ is 2300 pc \citep{djh01}.}
\tablenotetext{b}{Absorption corrected luminosity in the 2--10 keV band}
\end{center}
\end{table}

\end{document}